\documentclass[12pt,cqg]{iopart}
\pdfoutput=1

\usepackage{amsopn}
\usepackage{iopams}
\usepackage[numbers,square,sort&compress]{natbib}
\usepackage{graphicx}
\usepackage[pdftex]{hyperref}

\DeclareMathOperator*{\wlim}{w-lim}

\newcommand{\be}{\begin{equation}}
\newcommand{\ee}{\end{equation}}
\newcommand{\ba}{\begin{equation}\begin{aligned}}
\newcommand{\ea}{\end{aligned}\end{equation}}

\newcommand{\noteadded}[1]{\emph{\textbf{[NOTE ADDED:} #1\textbf{]}}}

\DeclareMathAlphabet{\mathpzc}{OT1}{pzc}{m}{it}

\begin{document}

\title{Comments on Backreaction}

\author{Stephen R. Green$^1$ and Robert M. Wald$^2$}

\address{$^1$ Perimeter Institute for Theoretical Physics \\ 31 Caroline Street North, Waterloo, Ontario N2L 2Y5, Canada}

\address{$^2$ Enrico Fermi Institute and Department of Physics, The University of Chicago\\5640 South Ellis Avenue, Chicago, Illinois 60637, USA}

\eads{\mailto{sgreen@perimeterinstitute.ca} and \mailto{rmwa@uchicago.edu}}

\begin{abstract} 
  We respond to the criticisms of a recent paper of Buchert et al.
\end{abstract}

\maketitle

Buchert et al have now published a revised
version~\cite{Buchert:2015ivb} of their original
paper~\cite{Buchert:2015iva} criticizing our work. In the revised
version, the tone of their criticisms has been improved and all
assertions that we made mathematical errors in our work have been
removed. However, the essential content of their paper remains
unchanged. Therefore, we shall hereby revise our rebuttal by simply
adding some remarks to our original rebuttal to note changes and
address new points. In addition,
elsewhere~\cite{Green:2015unpublished} we will give a simple,
heuristic discussion of why backreaction is negligible in
cosmology. Except for the remarks identified below as
``\noteadded{\ldots},'' the remainder of this paper is unchanged from
our original posting.

In a recent paper, Buchert et al~\cite{Buchert:2015iva} have
criticized nearly all aspects of our analysis of backreaction produced
by small scale inhomogeneities in cosmology as presented
in~\cite{Green:2010qy,Green:2011wc,Green:2013yua,Green:2014aga}. Most
of these criticisms concern points that are irrelevant to our main
arguments and/or are based upon misinterpretations of our
work. Nevertheless, in order to avoid the appearance that we are
ignoring or evading potentially valid issues, we will give a brief
response in the Appendix to all of the scientific/mathematical points
raised in~\cite{Buchert:2015iva}. In short, we stand firmly behind all
of the assertions, arguments, and conclusions presented in all of our
previous
papers~\cite{Green:2010qy,Green:2011wc,Green:2013yua,Green:2014aga}.

Nevertheless, we believe that the discussion of~\cite{Buchert:2015iva}
has the potential to create some confusion about the nature of our
analysis of backreaction and the conclusions that can be drawn from
it. Therefore, we feel that it would be appropriate for us to take
this opportunity to explain, in an informal way, what we mean by
``backreaction'' and how the main tools of our analysis fit in with
prior efforts, particularly the work of
Isaacson~\cite{Isaacson:1967zz,Isaacson:1968zza}.

The context for the phenomenon of backreaction by small scale
inhomogeneities concerns a situation where the actual spacetime metric
$g_{ab}$ has large curvature fluctuations on small scales but,
nevertheless, $g_{ab}$ can be well approximated by a metric
$g^{(0)}_{ab}$ that does not have large curvature
fluctuations. Although our analysis is valid in a much more general
context, the main situation we have in mind is where $g_{ab}$ is the
actual metric of the universe and $g^{(0)}_{ab}$ is a metric with FLRW
symmetry. For simplicity, we will restrict our discussion below to the
case where $g^{(0)}_{ab}$ has FLRW symmetry.

The issue at hand is whether the small scale inhomogeneities of
$g_{ab}$ can contribute nontrivially to the dynamics of
$g^{(0)}_{ab}$. A priori, this is possible even though
$\gamma_{ab} \equiv g_{ab} - g^{(0)}_{ab}$ is assumed to be small:
Einstein's equation for $g_{ab}$ contains \emph{derivatives} of
$\gamma_{ab}$, which need not be small even when $\gamma_{ab}$ is
small. Consequently, the Einstein tensor, $G_{ab}$, of $g_{ab}$ need
not be close to the Einstein tensor, $G^{(0)}_{ab}$, of
$g^{(0)}_{ab}$. Thus, although $g_{ab}$ is assumed to be an exact
solution of Einstein's equation with some stress-energy source
$T_{ab}$, it is possible that $g^{(0)}_{ab}$ may not be close to a
solution to Einstein's equation with a suitably averaged stress-energy
source $T^{(0)}_{ab}$. If this occurs, we say that there is a
substantial \emph{backreaction} effect of the small scale
inhomogeneities on the effective dynamics of $g^{(0)}_{ab}$.

There are two cases where our analysis does \emph{not} apply
(and was never intended or claimed to apply). The first case is where
the actual metric $g_{ab}$ is not close to any FLRW metric
$g^{(0)}_{ab}$. A good example of this relevant to cosmology is
obtained by taking $g_{ab}$ to be an LTB
model~\cite{Lemaitre:1933gd,Tolman:1934za,Bondi:1947fta,Moffat:1994qy,Alnes:2005rw}
with significant voids/overdensities on scales comparable to the
Hubble radius. A simpler (but much less physically motivated) example
discussed in~\cite{Ishibashi:2005sj} is the case where the universe
consists of two disconnected FLRW parts at different stages of
expansion/contraction. In such cases, since $g_{ab}$ differs
substantially from any single FLRW metric, there is little point in
even attempting to find a ``best fit'' FLRW metric $g^{(0)}_{ab}$ by
averaging or by any other scheme. If one insists on finding a ``best
fit'' $g^{(0)}_{ab}$, the dynamics of $g^{(0)}_{ab}$ would be expected
to differ significantly from standard FLRW dynamics. However, we would
not describe this as ``backreaction.''

The second case to which our analysis does not apply is where the
actual metric $g_{ab}$ \emph{is} close to an FLRW metric
$g^{(0)}_{ab}$, but one applies an averaging or other scheme to
$g_{ab}$ to construct an ``effective'' FLRW metric $g'^{(0)}_{ab}$
intended to reproduce averages of observables calculated in the actual
metric.  In general, this approach will give rise to a metric
$g'^{(0)}_{ab}$ that differs significantly from $g^{(0)}_{ab}$. This
will typically happen if one tries to match the geodesics of $g_{ab}$
in a naive way to corresponding geodesics in an FLRW model,
since---although $g_{ab}$ is close to $g^{(0)}_{ab}$---the geodesics
of $g_{ab}$ are not close to the geodesics of $g^{(0)}_{ab}$. We
provided an illustration of this in our ``ball bearing'' example
in~\cite{Green:2014aga}, where we constructed a smooth metric $q_{ab}$
on a $2$-sphere that was everywhere close to the round sphere metric
$q^{(0)}_{ab}$, but the geodesics and curvature of $q_{ab}$ are not
close to those of $q^{(0)}_{ab}$. As we explained
in~\cite{Green:2014aga}, observers living in ``Sphereland'' and doing
experiments with taut ropes (geodesics) would find extremely large
deviations from any round sphere model on small scales and also would
find some significant deviations on large scales. If they used their
geodesic and curvature data to determine a ``best fit'' round metric
$q'^{(0)}_{ab}$, they could easily make a substantial error, depending
upon exactly what ``averaging scheme'' they used. Returning to the
cosmological case, we noted in section 3 of~\cite{Green:2014aga} that
the Buchert approach uses timelike geodesics in an essential way and
the Clarkson and Umeh approach uses null geodesics in an essential
way, so \emph{a priori} both approaches may produce an FLRW metric
$g'^{(0)}_{ab}$ that differs significantly from $g^{(0)}_{ab}$. The
dynamics of $g'^{(0)}_{ab}$ may then differ significantly from that of
$g^{(0)}_{ab}$. If this occurs, we shall refer to the effect as
\emph{pseudo-backreaction}, since it has the appearance of being a
dynamical effect produced by small scale inhomogeneities but is
actually merely an artifact produced by a poor choice of
representative FLRW metric $g'^{(0)}_{ab}$. Our analysis does not
apply to pseudo-backreaction.

If, as we claim, naive averaging schemes making use of geodesics and
curvature are prone to producing errors in the choice of
representative FLRW model $g^{(0)}_{ab}$, how should one go about
determining $g^{(0)}_{ab}$? As we discussed in section 4.3
of~\cite{Green:2014aga}, this should be done in the manner that is
normally already done in practice by cosmologists (see, e.g., \cite{Ade:2015xua}):
Consider FLRW metrics (with appropriate free parameters) and their
perturbations (with appropriate free parameters). Calculate observable
quantities within the models and find the model parameters that
provide the best fit to all of the data. If one does not find an
acceptable fit, this shows that something is wrong or incomplete in
the theoretical model. If one does find an acceptable fit, this
determines the parameters of the model, and one thereby obtains a
representative FLRW metric $g^{(0)}_{ab}$. Of course, this does not
``prove'' that the theoretical model is correct or that $g^{(0)}_{ab}$
is close to the actual metric of the universe, $g_{ab}$. As in all
other areas of science, even if one has a simple model that provides
an extremely good fit to a wide variety of disparate data, one always
should be open to the possibility that there are alternative
explanations.

None of what has been said above addresses the issue of whether, in
our actual universe, there might be significant backreaction effects
on the large scale dynamics produced by small scale
inhomogeneities. As discussed above, this is possible \emph{a priori}
because even if $\gamma_{ab} = g_{ab} - g^{(0)}_{ab}$ is small, its
contribution to Einstein's equation need not be small. How can one
compute these backreaction effects?

The issue of backreaction was addressed in 1964 by Brill and
Hartle~\cite{Brill:1964zz} in the context of trying to find ``geon
solutions'' to Einstein's equation, consisting of a ball of
gravitational radiation that is held together by its
self-gravitation. In this case, the small scale inhomogeneities
consist of gravitational radiation, whose backreaction substantially
alters the background metric. The Brill-Hartle approach was then
significantly generalized by
Isaacson~\cite{Isaacson:1967zz,Isaacson:1968zza}. The Isaacson work is
nicely summarized in subsections 35.13--15 of Misner, Thorne, and
Wheeler~\cite{MTW}.

However, although the approximations made in Isaacson's work are well motivated
physically, some serious difficulties arise if one tries to give a
mathematically precise justification of them. The most serious
difficulties involve the justification of equations satisfied by
$\gamma_{ab}$ (denoted as $h_{\mu \nu}$ in~\cite{MTW}), such as
eq.~(35.59a) of~\cite{MTW}. The nature of these difficulties were
elucidated at the end of section III of our first
paper~\cite{Green:2010qy}.

In the late 1980's, one of us (R.M.W.) was very troubled by these
difficulties and felt that the best way to resolve them would be to
reformulate the Isaacson approximation scheme in terms of
one-parameter families of metrics $g_{ab}(\lambda)$, similar to what
is done to justify ordinary perturbation theory (see section 7.5
of~\cite{Wald:1984}). A reformulation of the Isaacson scheme in terms
of one-parameter families would allow one to say with much more
clarity what approximations are being made and would enable one to
rigorously derive the equations satisfied by the various quantities at
each order of approximation. However, to describe the Isaacson scheme
in this way, one would need to consider one parameter families where
$\gamma_{ab}(\lambda) \to 0$ as $\lambda \to 0$ but spacetime
derivatives of $\gamma_{ab}(\lambda)$ do not go to zero, in order that
that they may continue to make nontrivial contributions to the second
order Einstein tensor in the limit, thereby providing nontrivial
backreaction. It is therefore far from obvious how to describe the
limiting behavior as $\lambda \to 0$ in a mathematically precise
way. G. Burnett was a graduate student in the Chicago relativity group
and R.M.W. posed to Burnett the problem of reformulating the Isaacson
approximation scheme using one-parameter families. Burnett solved this
problem brilliantly in~\cite{Burnett:1989gp}, using the notion of weak
limits to give a mathematically precise characterization of the
one-parameter families $g_{ab}(\lambda)$. A discussion of the
relationship between Burnett's reformulation and the original Isaacson
scheme can be found in Burnett's paper and our previous papers
(particularly sections I and II of~\cite{Green:2010qy}).

Our contribution to the analysis of backreaction was to recognize that
the methods used in the Isaacson approach---as reformulated by
Burnett---could be directly imported to describe backreaction in
cosmology. The small scale inhomogeneities of primary interest now are
not gravitational waves oscillating in space and time but are the
perturbations associated with density inhomogeneities varying mainly
in space. Nevertheless, the scheme for calculating the ``Isaacson
average'' of the second order Einstein tensor of these perturbations
does not depend on any wavelike character of the perturbation and
works just as well for treating the backreaction effects of
Newtonian-like cosmological perturbations as it does for treating the
backreaction effects of gravitational waves.

Our general analysis presented in~\cite{Green:2010qy} allowed for the
presence of gravitational radiation. It thereby allowed for
backreaction produced by gravitational radiation as well as by matter
inhomogeneities. Our main results are contained in two theorems. These
theorems state that if the matter stress-energy tensor satisfies the
weak energy condition, then the effective stress-energy tensor
describing the leading order backreaction effects of the small scale
inhomogeneities must (i) be traceless and (ii) satisfy the weak energy
condition. In essence, our theorems say that significant backreaction
effects can be produced \emph{only} by gravitational radiation and not
by matter inhomogeneities.

At sub-leading order, matter inhomogeneities do produce nontrivial
backreaction effects, as we analyzed in~\cite{Green:2011wc}. In
particular, in Appendix B of~\cite{Green:2011wc} we computed the
backreaction effects of matter inhomogeneities on the expansion rate
of the universe, under the assumption that these inhomogeneities are
Newtonian-like in nature. A lengthy calculation revealed that
backreaction effectively modifies the matter stress-energy by adding
in the effects of kinetic motion and the Newtonian potential energy
and stresses. In particular, it ``renormalizes'' the proper mass
density to an ``ADM mass density.''

Our results are fully in accord with the ``back of the envelope''
estimates of backreaction given in~\cite{Ishibashi:2005sj}. Our
results on the sub-leading backreaction effects of Newtonian-like
matter perturbations are fully in accord with the analysis
of~\cite{Baumann:2010tm}.

\ack

This research was supported in part by NSF grants PHY-1202718 and
PHY-1505124 to the University of Chicago and by Perimeter Institute for Theoretical Physics. Research at
Perimeter Institute is supported by the Government of Canada through
Industry Canada and by the Province of Ontario through the Ministry of
Research and Innovation.

\appendix
\section{Response to Specific Criticisms} 

In this Appendix, we will provide brief responses to the specific
criticisms of~\cite{Buchert:2015iva}. However, we will not address the
following two categories of criticism: (1) Criticism or discussion of
any results not presented
in~\cite{Green:2010qy,Green:2011wc,Green:2013yua,Green:2014aga}; (2)
criticisms that are of no scientific importance. The second category
includes the subcategories of (a) criticisms that are unrelated to the
main issues at hand and (b) numerous insinuations and
innuendos\footnote{As a small but typical example, footnote 5
  of~\cite{Buchert:2015iva} states that ``Green and Wald obviously use
  a fixed $\lambda$-independent Riemannian metric for computing
  norms..." The word ``obviously'' suggests that we omitted to say
  this. In fact, we stated that we were using a fixed,
  $\lambda$-independent Riemannian metric for computing norms very
  clearly and explicitly in our papers. \noteadded{The word
  ``obviously'' has been deleted from footnote 5
  in~\cite{Buchert:2015ivb}.}} that suggest that we were
sloppy in our assertions and arguments and/or overlooked various
points. \noteadded{The tone has been improved
in~\cite{Buchert:2015ivb}.}  Our silence on these categories of
criticism should not be interpreted as indicating agreement. The
section numbers below correspond to the numbering
of~\cite{Buchert:2015iva}.

\medskip

\noindent \emph{Section 2}: This section criticizes our ball bearing
example (mentioned in the body of the paper above) where we
constructed a smooth metric $q_{ab}$ on a $2$-sphere that was
everywhere close to the round sphere metric $q^{(0)}_{ab}$, but the
geodesics and curvature of $q_{ab}$ are not close to those of
$q^{(0)}_{ab}$. The purpose of this example was simply to illustrate
how two metrics could be close without their geodesics and curvature
being close. Therefore, logically, it does not make sense to criticize
the example unless we made an error in our claim that the metrics are
close or our claim that their geodesics and curvature are not
close. Neither claim appears to be challenged in section 2. Thus, as a
matter of principle, we do not understand what is being criticized.

\medskip

\noindent \emph{Section 3}: This section points out that the Buchert
formalism violates the conclusions of our theorems on the effective
stress-energy tensor. The reasons why our results do not apply to the
Buchert formalism were already explained in the body of the paper
above. The remainder of this section mainly reviews our approach and
raises numerous mathematical questions concerning the validity of our
work, referring to Appendix B for details. We will give our response
to these points under ``Appendix B'' below, except for the following
comment: In contrast to what is said in subsection 3.7, backreaction
in our formalism does not suddenly ``turn on'' at $\lambda = 0$; the
one-parameter family of metrics $g_{ab}(\lambda)$ converges uniformly
on compact sets to $g^{(0)}_{ab}$, so the dynamics of $g^{(0)}_{ab}$
accurately describes the (large scale) dynamics of $g_{ab}(\lambda)$
for sufficiently small $\lambda$. Indeed, for $\lambda > 0$,
backreaction terms \emph{are} present, and are described to leading
order by the second order Einstein tensor; for $\lambda=0$, they
are accounted for by the effective stress-energy tensor
$t_{ab}^{(0)}$, which is the weak limit as $\lambda \to 0$ of the second order
Einstein tensor.

\medskip

\noindent \emph{Section 4}: Subsection 4.1 makes the point that
whenever one has a one-parameter family of metrics $g_{ab}(\lambda)$,
one could apply a $\lambda$-dependent diffeomorphism $\psi(\lambda)$
to generate a new one-parameter family of metrics
$\psi^*(\lambda) g_{ab}(\lambda)$. If $g_{ab}(\lambda)$ approaches a
limit as $\lambda \to 0$, then a $\psi(\lambda)$ may be chosen so that
$\psi^*(\lambda) g_{ab}(\lambda)$ approaches a different
limit\footnote{The fact that different limits can be obtained in this
  manner for spacetimes representing a body that ``shrinks to zero
  size'' plays a central role in the Gralla-Wald derivation of
  self-force~\cite{Gralla:2008fg,Gralla:2009md}.} or no limit at
all. In this sense, the limit of any one-parameter family of metrics
(or any other tensor fields) is always ``coordinate dependent.'' The situation with regard to our approach
is no different from ordinary perturbation theory (see section 7.5
of~\cite{Wald:1984}) where one could start with a one-parameter family
$g_{ab} (\lambda, x)$ that is jointly smooth in $\lambda$ and $x$, and then apply a diffeomorphism
$\psi(\lambda)$ that is not smooth in $\lambda$ at $\lambda = 0$ (as in the examples given in
subsection 4.1) so that $\psi^*(\lambda) g_{ab}(\lambda)$ approaches a different limit (or no limit) as $\lambda \to 0$. 
Obviously, the fact that one can perform such a construction is completely irrelevant to the validity of ordinary perturbation theory, and it is similarly irrelevant to the validity of our results.

A much less trivial issue---which does not appear to have been
considered in~\cite{Buchert:2015iva}---concerns the case where
$\psi^*(\lambda)$ goes to the identity as $\lambda \to 0$ (so that
$\wlim_{\lambda \to 0} \psi^*g_{ab} = g^{(0)}_{ab}$) but derivatives
of $\psi^*(\lambda)$ do not go to zero. Such transformations are of no
relevance if $\psi^*(\lambda) g_{ab}(\lambda)$ fails to satisfy our
conditions (i)--(iv), just as non-jointly-smooth diffeomorphisms are
of no relevance to ordinary perturbation theory. However, if
$\psi^*(\lambda) g_{ab}(\lambda)$ continues to satisfy our conditions
(i)--(iv), it gives rise to a gauge dependence of $\mu_{abcdef} \equiv
\wlim_{\lambda \to 0} \nabla_a \gamma_{bc} \nabla_d \gamma_{ef}$
analogous to the gauge dependence of quantities appearing in ordinary
perturbation theory. Nevertheless, as shown by
Burnett~\cite{Burnett:1989gp}, the effective stress-energy $t_{ab}$
describing backreaction is gauge invariant. This corresponds to the
statement that the Isaacson average of the second order Einstein
tensor becomes gauge invariant in the short-wavelength limit. Thus,
our backreaction results are fully ``coordinate invariant'' in the
desired sense.

In subsection 4.2, our example given in~\cite{Green:2013yua} of a
family of Gowdy metrics exhibiting nontrivial backreaction is
criticized on the grounds that it is ``ultra-local'' and also on the
grounds that we could have chosen a different family\footnote{Note
  that our choice of one-parameter family with nonvanishing
  backreaction gives rise to a $g_{ab}^{(0)}$ that provides a much
  better approximation to $g_{ab}(\lambda_0)$ than the alternative
  families suggested in subsection 4.2.} that would have given no
backreaction. Our purpose in giving the Gowdy family was to provide an
explicit example of a one-parameter family of metrics
$g_{ab}(\lambda)$ that satisfies our assumptions (i)--(iv) with
nonvanishing backreaction, $t_{ab} \neq 0$. Logically, it can be
invalidated only by showing that our choice of $g_{ab}(\lambda)$
violates at least one of our assumptions (i)--(iv) or that
$t_{ab} = 0$. This is not claimed/shown in subsection 4.2.

\medskip

\noindent \emph{Section 5}: This section is largely a defense against
our previous criticisms of the Buchert approach. We stand by our
previous criticisms. Subsection 5.4 discusses the issue of how to
obtain the ``smoothed'' metric $g^{(0)}_{ab}$ from $g_{ab}$.  As
already discussed above in the body of the paper, it is our view that
using geodesics and/or curvature of $g_{ab}$ to directly construct
$g^{(0)}_{ab}$ is a \emph{bad} way to proceed, since it is prone to
making significant errors because the geodesics and curvature of the
actual metric $g_{ab}$ differ significantly from
$g^{(0)}_{ab}$. Rather, one should determine $g^{(0)}_{ab}$ by
building a model for $g^{(0)}_{ab}$ and $\gamma_{ab}$ and fitting all
available data to the model.

\medskip

\noindent \emph{Section 6}: It is difficult for us to determine
exactly what is being criticized in this section, but we take this
opportunity to make two clarifying remarks related to the discussion:
(1) Our dictionary given in~\cite{Green:2011wc} that translates a
Newtonian model into a general relativistic spacetime will yield a
spacetime metric that solves Einstein's equation to very high
accuracy. The distinction between having a quantity nearly solve an
equation at all times versus having this quantity provide a good
approximation to a solution at all times was explained in the
second-to-last paragraph of the introduction
of~\cite{Green:2011wc}. (2) If $g_{ab}$ is close to $g^{(0)}_{ab}$,
then by flux conservation, the average apparent luminosity of sources
(including multiple images) in the metric $g_{ab}$ must closely match
that of $g^{(0)}_{ab}$. This is a simple fact, not an ``argument
against backreaction.'' \noteadded{In~\cite{Buchert:2015ivb} some text
  has been added to try to rebut this simple fact. Here the authors
  confuse average apparent luminosity (which closely matches that of
  the background FLRW metric, by conservation of flux) with average
  luminosity distance (which is a nonlinear function of apparent
  luminosity and thus can differ significantly from an FLRW
  model). The fact that nonlinear functions of average apparent
  luminosity can be significantly affected by inhomogeneities was
  explained clearly in section 4.3 of~\cite{Green:2014aga}.}

\medskip

\noindent \emph{Appendix B}: Appendix B contains the core of the
arguments against our results. It is referred to many times in the
body of~\cite{Buchert:2015iva} to justify their criticisms. Words such
as ``error'' appear frequently in Appendix B when describing of our
work. \noteadded{In~\cite{Buchert:2015ivb} these words have been
  removed.}

The authors of~\cite{Buchert:2015iva} appear to assume that we wished
to consider metrics $g_{ab}(\lambda)$ of low regularity in
spacetime. This is not the case. Indeed, until we
read~\cite{Buchert:2015iva}, it never occurred to us to consider
spacetime metrics that are not $C^\infty$ in spacetime at each fixed
$\lambda$, and there is no reason to consider metrics that are not
smooth in spacetime. \noteadded{In~\cite{Buchert:2015ivb}, the authors
  assert that ``one cannot safely carry out computations by declaring
  that second derivatives of the perturbing tensor $\gamma$ may be
  unboundedly large, and then treat them as ordinary smooth
  functions.'' This suggests that some of the misunderstandings of the
  authors may originate from a confusion between unboundedness of
  second spacetime derivatives in the limit $\lambda\to 0$ (which is
  allowed in our formalism) with unboundedness of these derivatives in
  spacetime at fixed $\lambda$ (which we had never contemplated
  allowing).} For smooth spacetime metrics, all of our manipulations
are trivially justified. For example, in their eq.~(B.9) one can
simply integrate by parts a second time to take all the derivatives
off of $\gamma_{ab}$, and then use our assumption (ii) (eq.~(7)
of~\cite{Buchert:2015iva}) to see that the limit vanishes.

Nevertheless, the discussion of Appendix B raises an issue of some
mathematical interest: If one did wish to consider metrics of low
spacetime regularity, precisely what regularity would be needed to
make our analysis valid? Indeed, since $\nabla_a \gamma_{bc}$ need not
have a pointwise limit as $\lambda \to 0$ and
$\nabla_a \nabla_b \gamma_{cd}$ may become unboundedly large as
$\lambda \to 0$, it would be surprising if our results actually
required smoothness of $\gamma_{ab}(\lambda)$ for all fixed
$\lambda > 0$.

To analyze the regularity needed, we note that our condition (ii)
(namely, $|\gamma_{ab}(\lambda, x)| < \lambda C_1(x)$) makes sense
only if $\gamma_{ab}$ is a tensor field that is defined (almost)
everywhere. Condition (ii) then immediately further implies that
$\gamma_{ab}$ is locally $L^1$ and thus defines a distribution. In
particular, its (weak) second derivatives\footnote{\noteadded{What we
    are calling the ``weak derivative'' here corresponds to what the
    authors of~\cite{Buchert:2015ivb} call the ``distributional
    derivative.''}} are automatically well defined as a distribution
via
\begin{equation}
\nabla_a \nabla_b \gamma_{cd} [f^{abcd}] \equiv \gamma_{cd} [\nabla_b \nabla_a f^{abcd}] = \int_M \gamma_{cd} \nabla_b \nabla_a f^{abcd}
\label{secder}
\end{equation}
for any test ($C^\infty_0$) tensor field $f^{abcd}$.  Similarly,
condition (iii) (namely,
$|\nabla_a \gamma_{bc}(\lambda, x)| < C_2(x)$) makes sense only if
$\gamma_{ab}$ is differentiable (almost) everywhere. Condition (iii)
then immediately further implies that $\nabla_a \gamma_{bc}$ is
locally $L^2$. It follows immediately from the above together with the
smoothness $g^{(0)}_{ab}$ that the Einstein tensor of
$g_{ab}(\lambda, x) = g^{(0)}_{ab}(x) +\gamma_{ab}(\lambda, x)$ is
well defined as a distribution on spacetime for all (sufficiently
small) $\lambda$. Thus, for all $\lambda > 0$, it makes sense to
require that Einstein's equation holds weakly. Our results then follow
by imposing Einstein's equation weakly at $\lambda > 0$ and taking the
weak limit as $\lambda \to 0$. The linear terms in the second
derivatives of $\gamma_{ab}$---the main focus of the critical remarks
of Appendix B---are easily seen to make no
contribution\footnote{\noteadded{Despite the improved tone of Appendix
    B, the fundamental misunderstandings displayed
    in~\cite{Buchert:2015iva} remain present
    in~\cite{Buchert:2015ivb}. For example, the first sentence of the
    paragraph following the paragraph containing eq.~(B.20)
    in~\cite{Buchert:2015ivb} asserts that
    ``$\wlim \nabla_a \nabla_b \gamma_{cd} = 0$ is \emph{not} a
    consequence of the GW hypotheses (ii), (iii), (iv) but a strong
    \emph{a priori} assumption \ldots'' This assertion is false, since
    $\wlim \nabla_a \nabla_b \gamma_{cd} = 0$ is a trivial consequence
    of~\eref{secder} and condition (ii).}} in this limit on account
of~\eref{secder} and condition (ii).

Thus, our results continue to hold under the weakest regularity
conditions under which our assumptions make any sense. This, of
course, is a side-point since we were, in any case, considering only
metrics that are smooth in spacetime at each fixed $\lambda$---in
which case all of our manipulations are much more trivially
justified. We conclude that the criticisms of Appendix B are
completely without merit. Similarly, all of the statements sprinkled
throughout the body of~\cite{Buchert:2015iva} that rely on Appendix B
for justification are completely without merit.

\medskip

\noindent \emph{Appendix C}: Appendix C purports to show that something
is wrong with our second example given in~\cite{Green:2013yua}. (Our
first example was criticized in subsection 4.2---see the discussion
above.) For our second example, we simply wrote down a one-parameter
family of metrics $g_{ab} (\lambda)$ that satisfied all of our
assumptions except Einstein's equation. We then declared it to solve
Einstein's equation with matter source
$T_{ab} (\lambda) = G_{ab} (\lambda)/8 \pi$ (``Synge's
method''~\cite{SyngesMethod}). As one might expect, the resulting
$T_{ab} (\lambda)$ does not satisfy the weak energy condition. We then
calculated the effective stress-energy of backreaction, $t_{ab}$, and
showed that it failed to satisfy the conclusions of our two theorems
(tracelessness and the weak energy condition). This example is useful
because it shows that, for the validity of our theorems, we cannot
dispense with the hypothesis of the weak energy condition on
$T_{ab} (\lambda)$.

The authors of~\cite{Buchert:2015iva} correctly say that we obtain our
matter stress-energy $T_{ab} (\lambda)$ by Synge's method, which they
call ``case (i).'' They then say that another approach would be to
obtain a different $T_{ab} (\lambda)$ from a conformally invariant
matter Lagrangian, which they call ``case (ii).'' Of course, case (ii)
has nothing to do with our work. In case (ii), the matter
stress-energy tensor would be conformally invariant, and their
eq.~(C.18) would hold. They then say, ``Let us look more closely at
Green and Wald's choice (i): using (C.18) \ldots'' They then proceed
to derive contradictions. But eq.~(C.18) applies only to case (ii),
not to case (i). \noteadded{In~\cite{Buchert:2015ivb}, the reference
  to eq.~(C.18) has been removed. Nevertheless, the association (made
  below eq.~(C.22)) of the weak limit of $T_{ab} (g(1))$ (which,
  obviously, is equal to $T_{ab} (g(1))$) with $T_{ab} (0)$ makes no
  sense to us.}

Since it is logically impossible for a metric and resulting
stress-energy tensor obtained by Synge's method to fail to be a
solution to Einstein's equation, it is logically impossible for our
example to be wrong unless we made a computational error in
calculating $t_{ab}$. We do not believe that we made any such error,
and no evidence of such an error is presented in Appendix C.

\bibliography{mybib}

\end{document}